%% file: main.tex
\documentclass[aps,twocolumn,pre, superscriptaddress]{revtex4-2}

\usepackage[english]{babel}
\usepackage{amsmath, amssymb, mathtools, physics}
\usepackage{braket}
\usepackage{ifthen}
\usepackage{graphicx,float}
\usepackage{multirow}
\usepackage[RGB,dvipsnames]{xcolor}

\usepackage{hyperref}
\usepackage[nameinlink,capitalise]{cleveref}
\usepackage{enumitem}
\usepackage{tikz}

\usepackage{pgfplots}
\renewcommand{\vec}{\boldsymbol}
\newcommand{\wtdpsi}[3]{\psi_{#1\to #2}(#3)}
\newcommand{\wtdpsieps}[4]{\psi^{#1}_{#2\to #3}(#4)}
\newcommand{\sigpw}{\sigma^\epsilon_{\text{pw}}}
\newcommand{\sigloc}{\sigma^\epsilon_{\text{loc}}}

\newcommand{\wtdnu}[3]{\nu_{#1\to #2}(#3)}
\newcommand{\wtdnueps}[4]{\nu^{#1}_{#2\to #3}(#4)}
\begin{document}
\title{Waiting-time based entropy estimators in continuous space without Markovian events}

\author{Jonas H. Fritz}

\author{Udo Seifert}

\affiliation{
 II. Institut für Theoretische Physik, Universität Stuttgart, 70550 Stuttgart, Germany
}
\date{\today}
\begin{abstract}
Estimating entropy production in continuous systems that can only be observed with a limited resolution remains an open problem in stochastic thermodynamics. Existing estimators based on the measurement of waiting-time distributions require either the detection of Markovian events, which uniquely determine the state of the system, or assume a discrete underlying dynamics. We present a novel estimator that relies solely on the detection of a single particle leaving or entering regions, or crossing manifolds, in continuous space. This estimator is based on the frequency and the duration of transitions between such events. We derive this bound by introducing two kinds of discretization of space. Finally, we compare our novel bound to the thermodynamic uncertainty relation improved by correlations using simulations of a Brownian vortex and discuss its relation to other lower bounds to entropy production.
\end{abstract}

\maketitle
\section{Introduction}
Stochastic thermodynamics often requires full access to the slow degrees of freedom of a system in order to determine quantities like heat exchange, work and entropy production \cite{seki10,jarz11,vdb15,peli21,shir23,seif25}. Inspired by experiments, which cannot resolve a microscopic system fully, thermodynamic inference attempts to infer underlying properties of the system from the measurement of thermodynamic quantities or to bound thermodynamic quantities based on such observations \cite{seif19,dieb25}. Lower \cite{seif25a} and upper \cite{nish23,bake23} bounds for the entropy production are particularly prominent, as this quantity measures the departure from equilibrium. 

Depending on the measurement scenario, these bounds are based on different observables, such as currents and their fluctuations \cite{bara15,ging16}, lumped states \cite{raha07,espo12,skin21,erte22}, correlation functions \cite{ober22,dech23,dieb25a} or waiting-time distributions between Markovian or renewal events \cite{vdm22,haru22,vdm22b}. Among these bounds, the last class has received special attention, since it performs better than the thermodynamic uncertainty relation (TUR) in most cases \cite{gu26}.

Still, challenges in applying these lower bounds on entropy production to experimental data remain \cite{song24,blom24,baie24}. Particularly for estimators based on waiting-time distributions, recent works have explored the impact of finite measurement statistics \cite{frit25,maie25} and erroneous \cite{vdm25} or stochastic \cite{maie25a} identification of renewal events. Even when it is error-prone, the observation of Markovian events, which determine the state of all slow degrees of freedom exactly, is required at some points along the trajectory. However, the prerequisite of having to measure Markovian events is quite restrictive, since they may not be accessible if the observation is a projection of a higher-dimensional dynamics \cite{gode22}. To rule out whether an event is Markovian, the measurement and quantification of memory has also received attention recently \cite{lapo21,voll24,zhao25}.

So far, only few estimators that omit the requirement of observing Markovian events and perform comparably have been found, one of which applies to blurred transitions in discrete systems \cite{erte24,haru24}, and another one which considers displacements in a fixed time interval in continuous systems \cite{lanz25}. Our approach generalizes the former by extending the results of Ref. \cite{erte24} to continuous systems and is conjugate to the latter. Instead of the distribution of displacements in a fixed time as in Ref. \cite{lanz25}, our bound considers the varying time transitions between fixed regions of space take. 

A previous attempt to extend the waiting-time based entropy estimator to continuous systems has been made \cite{meyb24}, but this approach relied on composite events consisting of two consecutive Markovian events in order to infer directionality and was only applicable to one-dimensional systems. Our approach is more general in that is also applicable to higher dimensional systems and does not rely on events with a finite duration. 

We start by introducing the dynamics and the type of observable our estimator is applicable to in \Cref{sec:setup}. We then present the estimator, highlight the conceptual difficulties in its proof and prove it in \Cref{sec:proof}. In \Cref{sec:numerics}, we demonstrate the bound using simulated data of a simplified Brownian gyrator \cite{fill07,dite24a} and compare the estimator to the correlation TUR (CTUR)\cite{piet17,horo17,dech21}. 

\section{Setup}
\label{sec:setup}
\begin{figure}
    \centering
    \includegraphics[width=0.9\linewidth]{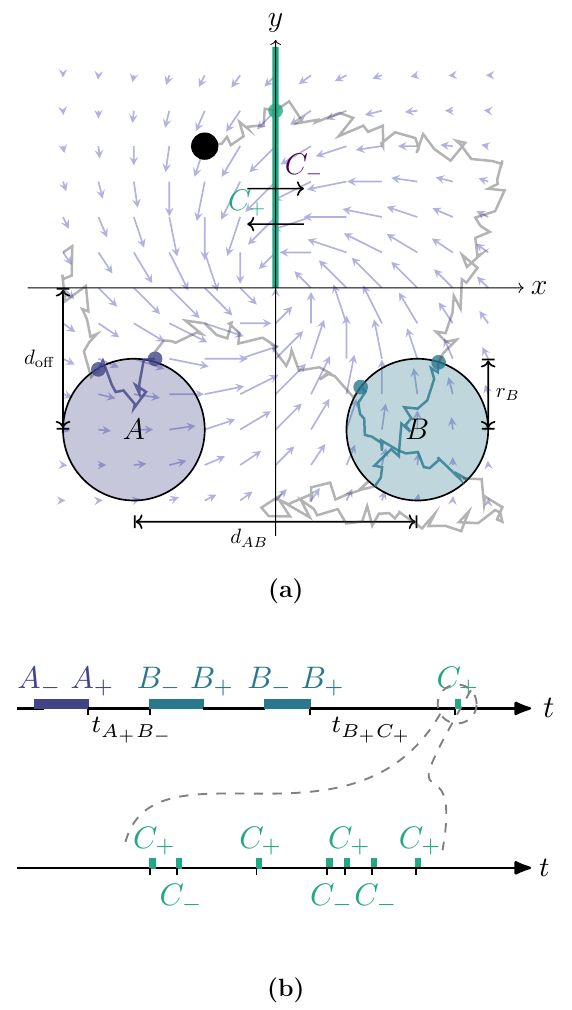}
    \caption{(a) Trajectory of an overdamped Langevin particle in two dimensions. The blue arrows show the steady state current at each point, given by the potential \cref{eq:num_pot} and non-conservative force \cref{eq:num_force}. Whenever the particle is in one of the colored areas, $A$ or $B$, or crosses the colored positive $y$-axis, the observer detects a signal which reveals which area the particle is in or in which direction it has crossed the positive $y$-axis, i.e., $C_+$ or $C_-$. (b) The resulting measurement. From such an infinitely long measurement, the frequency of transitions between $A_\pm,B_\pm$ and $C_\pm$ that take a certain time (like $t_{A_+ B_-}$) can be reconstructed. The enlarged timeline in the bottom illustrates the infinite number of recrossings that occur for any crossing of a boundary. The time $t_{A_+ B_-}$ ($t_{B_+ C_+}$) is then measured from the final occurrence of $A_+$ ($B_+$) to the first occurrence of $B_-$ ($C_+$). As we show in \cref{sec:scalings}, these recrossing events do not contribute to our entropy production estimator, but necessitate the introduction of two discretization schemes.}
    \label{fig:kartoffelwurm}
\end{figure}
\subsection{Dynamics}
We consider an overdamped Langevin dynamics in $d$ dimensions in which we allow for a space-dependent diffusion matrix $\vec D(\vec x)$. We focus on the steady state, in which the probability of observing the particle at $\vec x$, $\rho(\vec x)$, fulfills the Fokker-Planck equation \cite{risken,seif25}
\begin{equation}
    0=-\nabla\cdot \vec j(\vec x)\,,
\end{equation}
with the Cartesian components of the current
\begin{equation}
\begin{aligned}
    j_\alpha(\vec x)&=  \left[D_{\alpha\beta}(\vec x) F_\beta(\vec x) -D_{\alpha\beta}(\vec x)\partial_\beta \right]\rho(\vec x)\\
    &\equiv v_\alpha(\vec x)\rho(\vec x)\,,
    \end{aligned}
    \label{eq:def_curr}
\end{equation}
where we adopt the Einstein summation convention and the last equality defines the mean local velocity $\vec v(\vec x)$. The total force $\vec F(\vec x)$ can be split into two contributions
\begin{equation}
    \vec F(x)=\left[\vec f (\vec x)-\mathbf{\nabla} U(\vec x)\right]\,,
\end{equation}
where $U(\vec x)$ is an external potential and $\vec f(\vec x)$ denotes non-conservative forces. In the non-equilibrium steady state, the mean total entropy production rate of this system can then be written as 
\begin{equation}
\begin{aligned}
    \sigma &= \int d\vec x \,v_\alpha(\vec x)\left[D^{-1}(\vec x)\right]_{\alpha\beta}v_\beta(\vec x) \rho(\vec x)\\
    &= \langle v_\alpha(\vec x)\left[D^{-1}(\vec x)\right]_{\alpha\beta}v_\beta(\vec x) \rangle\,,
\end{aligned}
\label{eq:def_sig_nu_D_nu}
    \end{equation}
where $\langle\cdot \rangle$ indicates the steady state average.

On the trajectory level, the equation of motion reads
\begin{equation}
    \dot{x}^t_\alpha= D_{\alpha\gamma}(\vec x^t) F_\gamma(\vec x^t)+ g_{\alpha\gamma}(\vec x^t) \left[\partial_{\beta}g_{\beta\gamma}(\vec x^t)\right] + g_{\alpha\gamma}(\vec x^t)\zeta^t_\gamma\,,
    \label{eq:langevin_setup}
\end{equation}
where we set the inverse temperature $\beta =1$ and adopt the Stratonovich, or mid-point discretization rule. The noise is $\delta$-correlated
\begin{equation}
    \langle\vec\zeta^t_\alpha\vec\zeta_\beta^{t^\prime}\rangle = 2\delta_{\alpha\beta}\delta(t-t^\prime)\,,
\end{equation}
with the Kronecker symbol, $\delta_{\alpha\beta}$, and the diffusion matrix is connected to the noise amplitude $\vec{g}(\vec x)$ via
\begin{equation}
    D_{\alpha\beta}(\vec x) = g_{\alpha\gamma}(\vec x)g_{\beta\gamma}(\vec x)\,.
\end{equation}
The term $g_{\alpha\gamma}(\vec x^t) \left[\partial_{\beta}g_{\beta\gamma}(\vec x^t)\right]$ compensates for spurious drift in this convention.

\subsection{Observables}

We assume that an observer cannot fully resolve the trajectory $\vec x^t$. Rather, the particle is only registered when it is in certain $d$-dimensional regions in space or crosses certain $(d-1)$-dimensional manifolds, which we will often call boundaries. In particular, detecting the particle in certain regions is equivalent to detecting transitions into and out of the region through the region's enclosing boundary. Thus, we consider crossings of $(d-1)$-dimensional manifolds that may or may not enclose a $d$-dimensional subspace. While our arguments hold in $d$ dimensions, we will illustrate the concepts in a two-dimensional system for clarity. 

In the example shown in \Cref{fig:kartoffelwurm}(a), the particle can be detected whenever it is in either region $A$ or $B$, and whenever it crosses the positive $y$-axis in either direction. However, even if the particle is in either region, its precise location within that region cannot be measured. Similarly, while one might be able to resolve in which direction a crossing of the $y$-axis takes place, the precise location where it crosses is not accessible. Thus, at no point can an observer measure so-called Markovian events, which would allow for a factorization of the path weight in Markovian snippets and the estimator presented in \cite{vdm22b}. 

A measurement in which only the regions $A$ and $B$ can be observed might be achieved by single-molecule fluorescence correlation spectroscopy \cite{schw99} with multiple foci to differentiate the different regions \cite{otos19}.

The resulting observation is a sequence of colored signals and flashes, shown in \Cref{fig:kartoffelwurm} (b). A signal is detected whenever the particle is in one of the observable regions, with a dark waiting-time in between. Whenever the particle crosses a boundary such as the positive $y$-axis in this example, a corresponding flash is detected instantaneously. We denote an entry of a region $A$ by $A_-$ and the exit of this same region by $A_+$. Likewise, we assign the crossings of a boundary in the two possible directions $C_+$ and $C_-$, where the assignment is arbitrary but fixed. We now use the $\tilde{\cdot}$ symbol to indicate time reversal such that $\tilde{A}_\pm=A_\mp$ and $\tilde{C}_\pm=C_{\mp}$. If the directions of crossings across $C$ cannot be resolved, we denote the corresponding event by $C$ in which case we have $\tilde{C}=C$. Due to the white noise in \cref{eq:langevin_setup}, any initial crossing of a boundary will be followed by an infinite number of recrossings of this boundary. Eventually, the final crossing occurs after which the particle moves away from the boundary. We illustrate this for the event $C_+$ in the lower half of \cref{fig:kartoffelwurm} (b) and discuss it in more detail in \cref{sec:scalings}. 

By measuring e.g. the time between the final exit from $B$ to the first occurrence of $C_+$, we can count the number $n_{I\rightarrow J}(t_i)$ of transitions from an initial event $I$ to a final event $J$ that take a certain time $i\Delta t<t<(i+1)\Delta t$, with the bin width $\Delta t$. We can ignore the recrossing transitions $I\to\tilde{I}$, which we will explain in \cref{sec:scalings}. The frequency of transitions can be constructed from such an observation as
\begin{equation}
    \wtdnu{I}{J}{t} \equiv \lim_{T\rightarrow \infty} \lim_{\Delta t\rightarrow 0} \frac{n_{I\rightarrow J}(t_i)}{T\Delta t}\,.
    \label{eq:def_cap_nu}
\end{equation}
As we will show, this observable based on the occurrence of non-Markovian events, yields a lower bound on the mean total entropy production. 

Crucially, the type of observable we picture in this continuous setting implicitly assumes the properties of transition classes required in \cite{erte24}. Here, in a discrete system, transitions are lumped into transition classes, which have to be either odd or even under time reversal, i.e., under time-reversal, each transition class $I$ is mapped to either itself (even) or to another, unique transition class $\tilde{I}$, with $\tilde{\tilde{I}}=I$ (odd). 

In our case, the observation of states is equivalent to the observation of all exiting and entering transitions, where ``exiting" constitutes the time-reversed transition class of ``entering" and vice versa, such that these transition classes are odd under time-reversal. For crossing boundaries like the positive $y$-axis in the example above, transitions need to be observable across both directions, though not necessarily distinguishable. If the directions of transitions are not distinguishable, they form an even transition class. Else, they form an odd one. 

\section{Entropy estimator}
\label{sec:proof}
\subsection{Main result}
\begin{figure}
    \centering
    \includegraphics[width=0.9\linewidth]{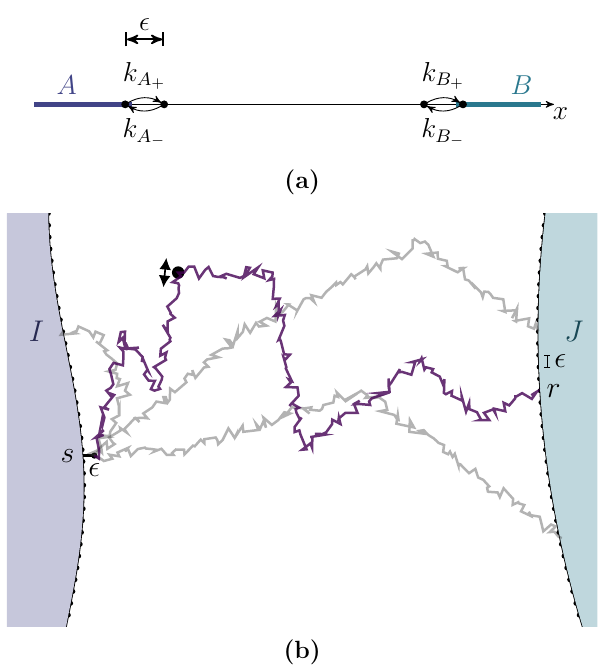}
    \caption{Two types of discretization. (a) Discretizing the distance to the boundary. In this one-dimensional model, the distance to the boundary needs to be discretized in order for the waiting-time distribution from $A$ to $B$ to be well defined. This scheme allows for the identification of $\nu^{\epsilon}_{A_+}$ as flux between the discretized states and the expansion of rates \cref{eq:approx_rates_1d}. (b) Discretizing the boundary itself. From this scheme, the scaling behavior of $\wtdpsieps{\epsilon}{i_s}{j_r}{t}$ as $\epsilon\to 0$ can be deduced. The probability of choosing a geometric, i.e., undirected path that reaches a specific segment $r$ on $J$ scales as $\epsilon^{d-1}$. Starting a distance $\epsilon$ away from the initial boundary, the particle can move forward and backwards along the specific geometric path marked in purple, as indicated by the black arrows. Thus, reabsorption at the initial boundary remains possible. The probability of reaching $J$ from $I$ given such a one-dimensional geometric path scales as $\epsilon$.}
    \label{fig:boundary}
\end{figure}
On a fundamental level, the entropy production rate measures the distance to equilibrium of a system by quantifying the time asymmetry as the log-ratio of the probability of a fully resolved forward trajectory $\gamma$ and a fully resolved backward trajectory $\tilde{\gamma}$. The steady state entropy production rate is then the average \cite{peli21,shir23,seif25}
\begin{equation}
    \sigma = \lim_{T\to\infty}\frac{1}{T}\left\langle\ln\frac{P\left[\gamma\right]}{\tilde{P}\left[\tilde{\gamma}\right]}\right\rangle\,.
\end{equation}
Typically, the full underlying trajectory cannot be resolved, but only some coarse-grained version $\Upsilon$, which has the path weight 
\begin{equation}
    P[\Upsilon] = \sum_{\gamma\in\Upsilon}P[\gamma]\,,
\end{equation}
where the sum is taken over all different paths $\gamma$ that appear as the coarse-grained path $\Upsilon$ to the observer. Crucially, this coarse-graining procedure needs to commute with time-reversal \cite{hart21a}, such that 
\begin{equation}
    \tilde{P}[\tilde{\Upsilon}] = \sum_{\tilde{\gamma}\in\tilde{\Upsilon}}\tilde{P}[\tilde{\gamma}]\,.
\end{equation}
For such a many-to-one mapping of microscopic trajectories $\gamma$ to observable trajectories $\Upsilon$, the log-sum inequality yields
\begin{equation}
    \lim_{T\to\infty}\frac{1}{T}\left\langle\ln\frac{P\left[\gamma\right]}{\tilde{P}\left[\tilde{\gamma}\right]}\right\rangle\geq \lim_{T\to\infty}\frac{1}{T}\left\langle\ln\frac{P\left[\Upsilon\right]}{\tilde{P}\left[\tilde{\Upsilon}\right]}\right\rangle\,.
    \label{eq:markov_logsum}
\end{equation}
Thus, the entropy production inferred from the coarse-grained trajectory is always less than that inferred from the fully resolved one. 

The trajectory $\Upsilon$ may feature certain events $i,j,\ldots$ that uniquely determine the state of the system, such that the path weight factorizes \cite{vdm22b} 
\begin{equation}
    P\left[\Gamma\right] = p(i_0)p(i_1,t_1|i_0,t_0)p(i_2,t_2|i_1,t_1)\ldots\,,
    \label{eq:markovian_factorization}
\end{equation}
where $\Gamma$ denotes a trajectory featuring such Markovian events. The probabilities 
\begin{equation}
    p(i_{n+1},t_{n+1}|i_{n},t_n)=p(i_{n+1},t_{n+1}-t_n|i_n)
\end{equation}
are called waiting-time distributions and depend only on the time difference $t_{n+1}-t_n$. For a more compact notation, we denote them by $\wtdpsi{i}{j}{t}$ for an arbitrary initial event $i$ and final event $j$. Integrating such a waiting-time distribution over time yields the transition probability
\begin{equation}
    p_{i\rightarrow j}  = \int_0^\infty\mathrm{d}t\wtdpsi{i}{j}{t}\,,
\end{equation}
and an additional summation over the final events yields the normalization
\begin{equation}
    \sum_{j}p_{i\rightarrow j}=1\,.
    \label{eq:normalization}
\end{equation}

Plugging the factorization \eqref{eq:markovian_factorization} in \cref{eq:markov_logsum} yields
\begin{equation}
\begin{aligned}
    \sigma\geq&\lim_{T\to\infty}\frac{1}{T} \left\langle\ln\frac{\wtdpsi{i_1}{i_2}{t_1}\ldots \wtdpsi{i_{n-1}}{i_n}{t_{n}}}{\wtdpsi{\tilde{i_n}}{\tilde{i}_{n-1}}{t_n}\ldots \wtdpsi{\tilde{i_2}}{\tilde{i_1}}{t_{n}}}\right\rangle\\
    =&\sum_{i,j}\int_0^\infty \mathrm{d}t\,\nu_{i}\wtdpsi{i_}{j}{t}\ln\frac{\wtdpsi{i}{j}{t}}{\wtdpsi{\tilde{j}}{\tilde{i}}{t}}\,,\\
\end{aligned}
\label{eq:markov_estimator}
\end{equation}
where the boundary terms vanish in the limit $T\to\infty$. For the second identification, we split the $\ln(\cdot)$ into a sum of corresponding time-reversed pairs $\ln[\wtdpsi{i}{j}{t}/\wtdpsi{\tilde{j}}{\tilde{i}}{t}]$. With the factor $1/T$, the number of terms with a specific triplet $i,j,t$ is the frequency of occurrence of such transitions, resulting in the second line of \cref{eq:markov_estimator}, with the frequency of the initial event $i$, $\nu_{i}$. Thus, waiting-time distributions between Markovian events provide a lower bound on the mean entropy production rate. This greatly reduces the size of the exponentially large space of possible trajectories that needs to be sampled to obtain an estimate of the entropy production rate, because the time between Markovian events is much shorter than the duration of the full trajectory. 

As a main result of this paper we show that a superficially similar bound based on the observable introduced in \cref{eq:def_cap_nu} also yields a lower bound on the entropy production rate of a system with the Langevin dynamics \eqref{eq:langevin_setup}
\begin{equation}
    \sigma\geq \sum_{I,J}\int_0^\infty \mathrm{d}t\, \wtdnu{I}{J}{t}\ln\frac{\wtdnu{I}{J}{t}}{\wtdnu{\tilde{J}}{\tilde{I}}{t}}\,,
    \label{eq:main_res}
\end{equation}
which, however, does not rely on the observation of Markovian events. In proving this bound, we will come across two conceptual challenges, one of which is best illustrated in a simple model in one dimension.

\subsection{The need for discretization}
\label{sec:scalings}
\subsubsection{Discretizing the distance to the manifold}
\label{sec:scalings_1}
We consider a simple one-dimensional model in which the particle moves in a confining potential and can be detected in two regions, one on the left and one on the right, see \Cref{fig:boundary} (a). Due to a lack of thermodynamic cycles, the steady state entropy production rate in this system is zero. Nevertheless, this system is useful to demonstrate one of the conceptual challenges in our proof. 

In this model, there are two relevant transition frequencies, $\nu_{A_+\to B_-}(t)$ and  $\nu_{B_+\to A_-}(t)$. These each split into two contributions
\begin{equation}
    \wtdnu{I}{J}{t}= \nu_{I}\wtdpsi{I}{J}{t}\,,
    \label{eq:factorization}
\end{equation}
with $I\in\{A_+,B_+\}$ and $J\in\{A_-,B_-\}$. The frequency with which the particle exits $I$ is $\nu_{I}$ and $\wtdpsi{I}{J}{t}$ is the waiting-time distribution of entering $J$ after time $t$, given that the particle has left the region $I$ at time $0$. 

The first conceptual difficulty we face is that the quantities $\nu_{I}$ and $\wtdpsi{I}{J}{t}$ are formally ill-defined. Let us consider, for example, a trajectory in which the particle leaves $A$ and enters $B$ some time later. After exiting $A$ and starting at a point $x_{A}$ at time $t=0$, the particle will recross this point infinitely often in a finite time interval before reaching the boundary of $B$ at $x_B$. Even though there is a final point in time at which the particle is at $x_A$, the particle will be at $x_A$ infinitely often between $t=0$ and this final time. Thus, the event $A_+$ occurs infinitely often before the event $B_-$ occurs, making $\nu_{A_+}$ appear to be infinite. Additionally, the transition probability appears to be $p_{A_+\rightarrow J}=\delta_{A_+,A_-}$, which would lead to the false conclusion that transitions from the region $A$ to the region $B$ do not happen. 

Further, the waiting-time distribution $\wtdpsi{A_+}{B_-}{t}$ is obtained by solving the Fokker-Planck equation corresponding to \cref{eq:langevin_setup} with absorbing boundary conditions at the observable states and initial condition $p(x,t=0)=\delta(x- x_A)$. Since $x_A$ is itself an observable point and thus an absorbing boundary, the initial condition places the particle on an absorbing boundary. 

Both of these apparent contradictions can be resolved by discretizing the space around $x_A$ \cite{maju05,bark14,mori19}, thus letting the particle start a distance $\epsilon$ away from the boundary. This scheme is illustrated in \Cref{fig:boundary} (a). At each boundary, there are two discrete states: one that sits just within the observable region, $x_{A,B}$, and one that is a distance $\epsilon$ away from it, $x_{A,B}\pm\epsilon$. This discretization allows us to calculate how the frequency of $A_+$ in this discretized system, $\nu^{\epsilon}_{A_+}$, diverges as $\epsilon\to 0$ by identifying $\nu^{\epsilon}_{A_+}$ as the flux across the observable boundary
\begin{equation}
    \nu^{\epsilon}_{A_+} = p_{x_A}k_{A_+}\,,
\end{equation}
where $p_{x_A}$ is the probability of finding the particle at the discretized state $x_A$. Both the forward rate $k_{A_+}$ and the backward rate $k_{A_-}$ can be expanded as \cite{piet17b}
\begin{equation}
    k_{A_{\pm}}\approx\frac{D( x_A)}{\epsilon^2}\left(1\pm\frac{f(x_A)\epsilon}{2}\right)\left(1\mp\epsilon\frac{\partial_x V(x_A)}{2}\right)
    \label{eq:approx_rates_1d}
\end{equation}
where the leading order diffusive contribution scales as $\epsilon^{-2}$. As a one-dimensional probability, $p_{x_A}$ scales with $\epsilon$, such that overall, $\nu^{\epsilon}_{A_+}$ diverges with $1/\epsilon$.

Placing the particle a distance $\epsilon$ away from the closest absorbing boundary also leads to a well defined waiting-time distribution, for which we can then analyze the $\epsilon\to 0$ behavior. 
The waiting-time distribution can be formally split into two probabilities
\begin{equation}
\begin{aligned}
    \wtdpsieps{\epsilon}{A_+}{B_-}{t} &= p^{\epsilon}(\text{no return})p^{\epsilon}(B_-,t|A_+,\text{no return})\,.\\
\end{aligned}
    \label{eq:psi_scaling_1d}
\end{equation}
Having started from $x_A+\epsilon$, $p^{\epsilon}(\text{no return})$ is the probability of not returning to $x_A$ before reaching $x_B$ and $p^{\epsilon}(B_-,t|A_+,\text{no return})$ is the probability of reaching $x_B$ at time $t$ \textit{conditional} to not returning to $x_A$ before. The latter is of order $\epsilon^0=\mathcal{O}(1)$, because in one dimension, the particle is sure to reach $x_B$ eventually if it is not absorbed at $x_A$. The former is the so-called splitting probability, for which no general solution exists. However, for a constant force $F(x)=f$, and after choosing coordinates in which $x_A=0$, the probability of reaching $x_B$ starting from $x(t=0)=\epsilon$ can be explicitly calculated \cite{redn01}
\begin{equation}
    p^{\epsilon}(\text{no return}) = \frac{1-e^{-f\epsilon }}{1-e^{-fx_B}}=\frac{f}{1-e^{-fx_B}}\epsilon+\mathcal{O}(\epsilon^2)\,.
    \label{eq:splitting_1d}
\end{equation}

With this result, we argue that the scaling $p^{\epsilon}(\text{no return})\sim\epsilon$ is universal in leading order. For any smooth total force $F(x)$, we can choose an intermediate point $x_I$ that is close enough to $x_A$ such that a constant force between $x_A$ and $x_I$ is a good approximation. Then, we can split the splitting probability as
\begin{equation}
    p^{\epsilon}(\text{no return})= p^{\epsilon}(x_I|x_A)p(x_B|x_I),
\end{equation}
where $p^{\epsilon}(x_I|x_A)$ is the probability of reaching $x_I$ starting from $x_A+\epsilon$ without returning to $x_A$ and $p(x_B|x_I)$ is the probability of reaching $x_B$ starting from $x_I$ without returning to $x_A$. Since the latter is independent of $\epsilon$ and the former is equal to \cref{eq:splitting_1d} after replacing $x_B$ with $x_I$, $\epsilon^1$ is always the leading order in the scaling of $p^{\epsilon}(\text{no return})$. Thus, $\wtdpsieps{\epsilon}{A_+}{B_-}{t}$ vanishes with order $\epsilon$ in leading order in one dimension.

Overall, with $\nu^{\epsilon}_{A_+}$ diverging with $1/\epsilon$ and $\wtdpsieps{\epsilon}{A_+}{B_-}{t}$ vanishing with $\epsilon$, their product $\wtdnueps{\epsilon}{A_+}{B_-}{t}$ remains a well-defined quantity in one dimension. 

We have thus established the need to discretize the space around the boundary of observable regions and observable boundaries in one dimension. Since the apparent contradictions we have illustrated in one dimension also occur in higher dimensions, the same discretization needs to be performed. For any point on the $(d-1)$-dimensional manifold, $\vec x_s$, we need to introduce another discrete state close to the manifold, $\vec x_s+\epsilon \vec n$, where $\vec n$ is the normal vector in the direction of the transition in question. In \cref{fig:kartoffelwurm} (a), these states would sit to the left (right) of the positive $y$-axis for the initial transition $C_+$ ($C_-$). Moreover, in higher dimensions, a further discretization is also necessary.

\subsubsection{Discretizing the manifold itself}
For our proof, we need to consider a more detailed level of description, in which the exit point $s$ along the boundary $\partial I$ and the entry point $r$ on $\partial J$ can be resolved. We can then make use of the fact that these events are Markovian, which after some additional algebra lets us apply the bound \cref{eq:markov_estimator}. These detailed entry and exit events are denoted by $i_s$ and $j_r$, respectively. 

As we have established, the particle must be placed a distance $\epsilon$ away from the boundary for the frequency of the initial event $\nu^{\epsilon}_{i_s}$ and the waiting-time distribution from $i_s$ to $j_r$ to be well-defined. In one dimension, this was sufficient, as there can be at most a countable number of enclosing points for any observable region. In higher dimensions, this is no longer the case, as we illustrate in \Cref{fig:boundary}(b). The number of points on the enclosing boundary of $I$ and $J$ is infinite, such that the probability of hitting the $J$ at precisely the point $r$ is zero. 

Again, the solution is to introduce a second, conceptually different discretization: rather than considering points on the boundary, we consider small segments of size $\epsilon$. Then, every transition $i_s\to j_r$ from a small segment $s$ on $I$ to another small segment $r$ on $J$ contributes to the coarse-grained transition from $I$ to $J$, such that the coarse-grained frequencies are connected to the fine-grained ones via
\begin{equation}
    \wtdnu{I}{J}{t} = \sum_{s\in\partial I}\sum_{r\in\partial J}\,  \wtdnueps{\epsilon}{i_s}{j_r}{t} =\sum_{s\in\partial I}\sum_{r\in\partial J}\nu^{\epsilon}_{i_s}\wtdpsieps{\epsilon}{i_s}{j_r}{t} \,.
    \label{eq:nu_cap_from_nu_small}
\end{equation}
For a finite segment size, the events $i_s$ are no longer Markovian, since the state of the particle is no longer uniquely determined by the occurrence of $i_s$. However, Markovianity is recovered in the limit $\epsilon\to 0$. Therefore, we are interested in the scaling of the quantities on the right-hand side of \cref{eq:nu_cap_from_nu_small} in this limit. We deduce this scaling for each quantity in $d$ dimensions with the following simple arguments.

The expansion \cref{eq:approx_rates_1d} holds regardless of $d$, such that the jump rate scales as $\epsilon^{-2}$ in all dimensions. The probability of observing the particle at the discretized state within the region, however, scales with $\epsilon^d$. Thus, $\nu^{\epsilon}_{i_s}$ scales as $\epsilon^{d-2}$. 

For $\wtdpsieps{\epsilon}{i_s}{j_r}{t}$, we consider all geometric, i.e., undirected paths between the initial point $\vec x_s+\epsilon\vec n$ that the particle is at after the event $i_s$ has occurred and some final point $\vec y_r$ that the particle is at after the final absorbing event $j_r$ has occurred. In $d$ dimensions, the absorbing boundaries are $(d-1)$-dimensional manifolds. Thus, the probability of selecting a geometric path to any particular segment $r$ scales as $\epsilon^{d-1}$. However, this probability is conditional to not returning to the initial point, as in \cref{eq:psi_scaling_1d}. We illustrate this in \Cref{fig:boundary} (b), where we show some of the possible paths to other segments in gray. Along a particular geometric path, marked in purple, the particle may take forward and backwards steps and may even be reabsorbed at $\vec x_s$. Thus, we also need to consider the probability of successfully completing the geometric path from $\vec x_s$ to $\vec y_r$. As each geometric path is one-dimensional, regardless of how many dimensions it is embedded in, we can employ the arguments from \cref{sec:scalings_1} for the probability of successfully completing the transition. This probability then scales with $\epsilon^1$ in leading order. Thus, in total $\wtdpsieps{\epsilon}{i_s}{j_r}{t}$ scales as $\epsilon^{d}$ and thus $\wtdnueps{\epsilon}{i_s}{j_r}{t}$ scales as $\epsilon^{2(d-1)}$. Note that this is the frequency with which the transitions between vanishingly small segments of $(d-1)$-dimensional manifolds occur which is expected to vanish. When considering the corse-grained observable in \cref{eq:nu_cap_from_nu_small}, we still need to perform a double summation over the initial and final microscopic events. The number of summands in both $\sum_{s\in\partial I}$ and $\sum_{r\in\partial J}$ scales with $\epsilon^{-(d-1)}$, such that expression on the left-hand side of \cref{eq:nu_cap_from_nu_small} remains well-defined of $\mathcal{O}(1)$. 

To summarize, the different scaling behaviors are
\begin{align}
    p_i&\sim\epsilon^d\,,\\
    k_{i_s}&\sim \epsilon^{-2}\,,\\
    \nu^{\epsilon}_{i_s}&\sim \epsilon^{d-2}\,,\\
    p^{\epsilon}(\text{no return})&\sim \epsilon\,,\\
    \wtdpsieps{\epsilon}{i_s}{j_r}{t}&\sim\epsilon^d\,,\\
        \label{eq:psi_scaling}
    \sum_{s\in\partial I}&\sim \epsilon^{-(d-1)}\,,\\
    \text{and}\quad\wtdnu{I}{J}{t}&\sim \epsilon^0=\mathcal{O}(1)\,.
\end{align}

Fluctuations of empirical density and current are other observables for which the need to coarse-grain continuous space has been established \cite{dieb22}. With these preliminary remarks, we can move on to prove the bound \cref{eq:main_res}.

\subsection{Proving the bound}
We start by applying the log-sum inequality to \cref{eq:main_res}
\begin{widetext}
\begin{equation}
    \sum_{I,J}\int_0^\infty \mathrm{d}t\, \wtdnu{I}{J}{t}\ln\frac{\wtdnu{I}{J}{t}}{\wtdnu{\tilde{J}}{\tilde{I}}{t}}\leq \sum_{I,J}\sum_{\substack{s\in \partial I\\r\in\partial J}}\int_0^\infty \mathrm{d}t\, \wtdnueps{\epsilon}{i_s}{j_r}{t}\ln\frac{\wtdnueps{\epsilon}{i_s}{j_r}{t}}{\wtdnueps{\epsilon}{\tilde{j_r}}{\tilde{i}_s}{t}}\,,
    \label{eq:log_sum}
\end{equation}
\end{widetext}
which holds for any small $\epsilon$ and $I,J$ run over all events including their time reversed ones. Note that in the special case of transitions $I_\pm\to I_\mp$ on the left-hand side and transitions $i_s\to\tilde{i_s}$ on the right-hand side, the $\ln(\cdot)$ term is identical to $0$, since after applying the time reversal, the nominator and denominator are the same. Thus, the diverging frequency of recrossing transitions is not an issue in this sum. Similarly, the frequency of transitions which cannot occur due to the topology of the regions and boundaries, like $A_-\to A_-$, is zero, such that the corresponding terms do not contribute to the sums on either side of inequality \eqref{eq:log_sum}. Applying the factorization \eqref{eq:factorization} to the right-hand side of inequality \eqref{eq:log_sum} leads to 
\begin{equation}
\begin{aligned}
    &\sum_{I, J}\sum_{\substack{s\in\partial I\\r\in\partial J}}\int_0^\infty \mathrm{d}t\, \wtdnueps{\epsilon}{i_s}{j_r}{t}\ln\frac{\wtdnueps{\epsilon}{i_s}{j_r}{t}}{\wtdnueps{\epsilon}{\tilde{j_r}}{\tilde{i_s}}{t}}\\
    =&\sum_{I, J}\sum_{\substack{s\in\partial I\\r\in\partial J}}\int_0^\infty \mathrm{d}t\,\nu^{\epsilon}_{i_s}\wtdpsieps{\epsilon}{i_s}{j_r}{t}\ln\frac{\wtdpsieps{\epsilon}{i_s}{j_r}{t}}{\wtdpsieps{\epsilon}{\tilde{j_r}}{\tilde{i_s}}{t}}\\
    &+\sum_{I, J}\sum_{\substack{s\in\partial I\\r\in\partial J}}\nu^{\epsilon}_{i_s}p^{\epsilon}_{i_s\rightarrow j_r}\ln\frac{\nu^{\epsilon}_{i_s}}{\nu^{\epsilon}_{\tilde{j_r}}}\\
    \equiv & \sigpw+\sigloc\,,
    \end{aligned}
    \label{eq:split_contrib}
\end{equation}

In the following, we will show that $\sigpw$, which is related to the path weight of a trajectory $\Gamma$, provides a lower bound to $\sigma$ in the limit $\epsilon\to 0$. Moreover, $\sigloc$ is the local entropy production of the observable boundaries and will be shown to vanish. 

As in \cref{eq:markov_estimator}, $\sigpw$ can be rewritten as
\begin{equation}
    \sigpw = \lim_{T\to\infty}\frac{1}{T}\left\langle\ln\frac{\wtdpsieps{\epsilon}{i_0}{i_1}{t_1}\wtdpsieps{\epsilon}{i_1}{i_2}{t_2}\ldots}{\wtdpsieps{\epsilon}{\tilde{i_1}}{\tilde{i_0}}{t_1}\wtdpsieps{\epsilon}{\tilde{i_2}}{\tilde{i_1}}{t_2}\ldots}\right\rangle\,,
    \label{eq:loglim_sigpw}
\end{equation}
where on the right-hand side, $i_0,i_1,\ldots$ is the sequence of events in a long trajectory, without recurrent events $i_s\to\tilde{i_s}$. In order to later identify the nominator and denominator as path weights of the forward and backward trajectory, respectively, these recurrent events can be added, since the corresponding log ratio is zero. Further, we need to take the limit $\epsilon\to0$, such that each event $i_s$ becomes Markovian. We thus recover a factorization of the path weight as in \cref{eq:markovian_factorization}. Keeping track of the orders of $\epsilon$ using the scaling we have derived in \cref{sec:scalings} yields
\begin{equation}
\begin{aligned}
    \wtdpsieps{\epsilon}{i_1}{i_2}{t_1}\times\ldots\times \wtdpsieps{\epsilon}{i_{n}}{i_{n+1}}{t_{n}}=\mathcal{O}(\epsilon^{nd})\,.
    \end{aligned}
\end{equation}
The same scaling holds for the denominator, as it contains the same number of waiting-time distributions, such that the dependence on $\epsilon$ in \cref{eq:loglim_sigpw} cancels. Thus, the expression $\sigpw$ stays well-defined in the limit $\epsilon\to 0$ and can be interpreted as the bound to entropy production \cref{eq:markov_estimator}
\begin{equation}
   \lim_{\epsilon\to0}\sigpw\leq\sigma\,,
\end{equation}
which completes the first half of our proof.

By using the normalization \cref{eq:normalization}, we can rewrite the second term on the right-hand side of \cref{eq:split_contrib} as 
\begin{equation}
    \sigloc=\sum_{I, J}\sum_{\substack{s\in\partial I\\r\in\partial J}}\nu^{\epsilon}_{i_s}p^{\epsilon}_{i_s\rightarrow j_r}\ln\frac{\nu^{\epsilon}_{i_s}}{\nu^{\epsilon}_{\tilde{j_r}}} = \sum_I\sum_{s\in \partial I}(\nu^{\epsilon}_{i_s}-\nu^{\epsilon}_{\tilde{i_s}})\ln\frac{\nu^{\epsilon}_{i_s}}{\nu^{\epsilon}_{\tilde{i_s}}}\,,
    \label{eq:sig_rt}
\end{equation}
where the summation runs over each segment once. By once again introducing a discretization in the vicinity of the point $\vec{x}_s$, as in \Cref{fig:boundary}(a), we can identify
\begin{equation}
    \nu^{\epsilon}_{i_s}=k^+(\vec x_s)p(\vec{x}_s)\,;\quad \nu^{\epsilon}_{\tilde{i_s}} = k^-(\vec x_s)p(\vec{x_s}+\epsilon\vec n)\,,
\end{equation}
where $p(\vec x_s)=\epsilon^d\rho(\vec x_s)$ is the probability on the discretized network in $d$ dimensions, $k^+$ is the jump rate for the transition corresponding to the occurrence of $i_s$ and $k^-$ the one for $\tilde{i_s}$. We can once again use the expansion \cref{eq:approx_rates_1d} by using $\vec n\vec D(\vec x_s)\vec n$ as the local diffusion coefficient, and by replacing the forces by their projection along $\vec n$. By also expanding the probabilities as 
\begin{equation}
    p(\vec x_s+\epsilon\vec n)\approx p(\vec x_s)(1+\epsilon\vec n\nabla_n\ln p(\vec x_s))\,,
\end{equation}
the log-ratio in \cref{eq:sig_rt} becomes
\begin{equation}
\begin{aligned}
       \ln\frac{\nu^{\epsilon}_{i_s}}{\nu^{\epsilon}_{\tilde{i_s}}}=& \ln\frac{p(\vec x_s) k^+(\vec x_s)}{p(\vec x_s+\epsilon\vec n))k^-(\vec x_s)}\\
        \approx& \epsilon \left[F_n(\vec x_s) -\nabla_n \ln p(\vec x_s)\right]=\epsilon\frac{v_n(\vec x_s)}{\vec n\vec D(\vec x_s)\vec n}\,,
\end{aligned}
\label{eq:approx_log}
\end{equation}
in leading order of $\epsilon$ and by identifying the local mean velocity from \cref{eq:def_curr}. 

Through the same expansions we obtain
\begin{equation}
\begin{aligned}
    \nu^{\epsilon}_{i_s}-\nu^{\epsilon}_{\tilde{i_s}}\approx&  \frac{1}{\epsilon}p(\vec x_s)\left(\vec n \vec D(\vec x_s)\vec n\right)\left(F_n(\vec x_s)-\vec n\nabla_n\ln p(\vec x_s)\right)\\
    =&\frac{p(\vec x_s)}{\epsilon} v_n (\vec x_s)\,,
    \end{aligned}
    \label{eq:approx_curr}
\end{equation}
where we again identify the mean local velocity from \cref{eq:def_curr}. Plugging \cref{eq:approx_log} and \cref{eq:approx_curr} in \cref{eq:sig_rt} yields
\begin{equation}
    \sum_I\sum_{s\in \partial I}(\nu^{\epsilon}_{i_s}-\nu^{\epsilon}_{\tilde{i_s}})\ln\frac{\nu^{\epsilon}_{i_s}}{\nu^{\epsilon}_{\tilde{i_s}}} = \sum_{x_s}p(\vec x_s)\frac{v^2_n(\vec x_s)}{\vec n \vec D(\vec x_s)\vec n}\,.
\end{equation}
Finally, we replace the sum over probabilities with an integral over the probability density. Taking into account that only the transitions across any observable boundary are summed over, the summation turns into an integral across the close proximity of the boundary
\begin{equation}
    \sum_{x_s}p(\vec x_s)\rightarrow \sum_I\int_{\partial I}\mathrm{d}s\int_{0}^{\epsilon} \mathrm{d}n\, ,
\end{equation}
where the first integral is a line integral across each observable $(d-1)$-dimensional manifold and the second integral is over the direction perpendicular to this surface. We can then further identify the local entropy production rate from \cref{eq:def_sig_nu_D_nu} 
\begin{equation}
\begin{aligned}
    &\sum_I\int_{\partial I}\mathrm{d}s\int_{0}^{\epsilon} \mathrm{d}n\, \rho(\vec x_s)\frac{v^2_n(\vec x_s)}{\vec n \vec D(\vec x_s)\vec n}\\
    =&\sum_I\int_{\partial I}\mathrm{d}s\int_{0}^{\epsilon} \mathrm{d}n\,\rho(\vec x_s)\sigma(\vec x_s)\,,
\end{aligned}
    \label{eq:second_term_lim}
\end{equation}
which establishes the relation between $\sigloc$ and the local entropy production rate in \cref{eq:def_sig_nu_D_nu}. Since this local entropy production rate is integrated only over the observable $(d-1)$-dimensional manifolds, the integral in \cref{eq:second_term_lim} vanishes in the limit $\epsilon\to0$ for non-singular forces, and we end up with the inequality \cref{eq:main_res}.

In contrast, in Ref. \cite{erte24} a similar term in Markov networks yields a finite contribution to the entropy bound even for non-singular forces. As we have shown, the expression in \cref{eq:sig_rt} can be seen as a local entropy production rate. In Markov networks, this is the entropy production associated with transitions across observable links. Except when all observable links fulfill local detailed balance, this yields a finite contribution and may even be the full entropy production of the Markov network, if only observable links break detailed balance. In continuous systems, the expression \cref{eq:sig_rt} is the entropy production associated with the displacement over a boundary. Since the distance between either side of a boundary vanishes, so does the entropy production associated with this displacement. Only in the case of singular forces that would arise from a potential step, this contribution is of $\mathcal{O}(1)$. It is then still bounded by $\sigma$, which would introduce an additional factor $1/2$ to the estimator \cref{eq:main_res}. 

\section{Numerical demonstration}
\label{sec:numerics}
\begin{figure*}
    \centering
    \includegraphics[width=\linewidth]{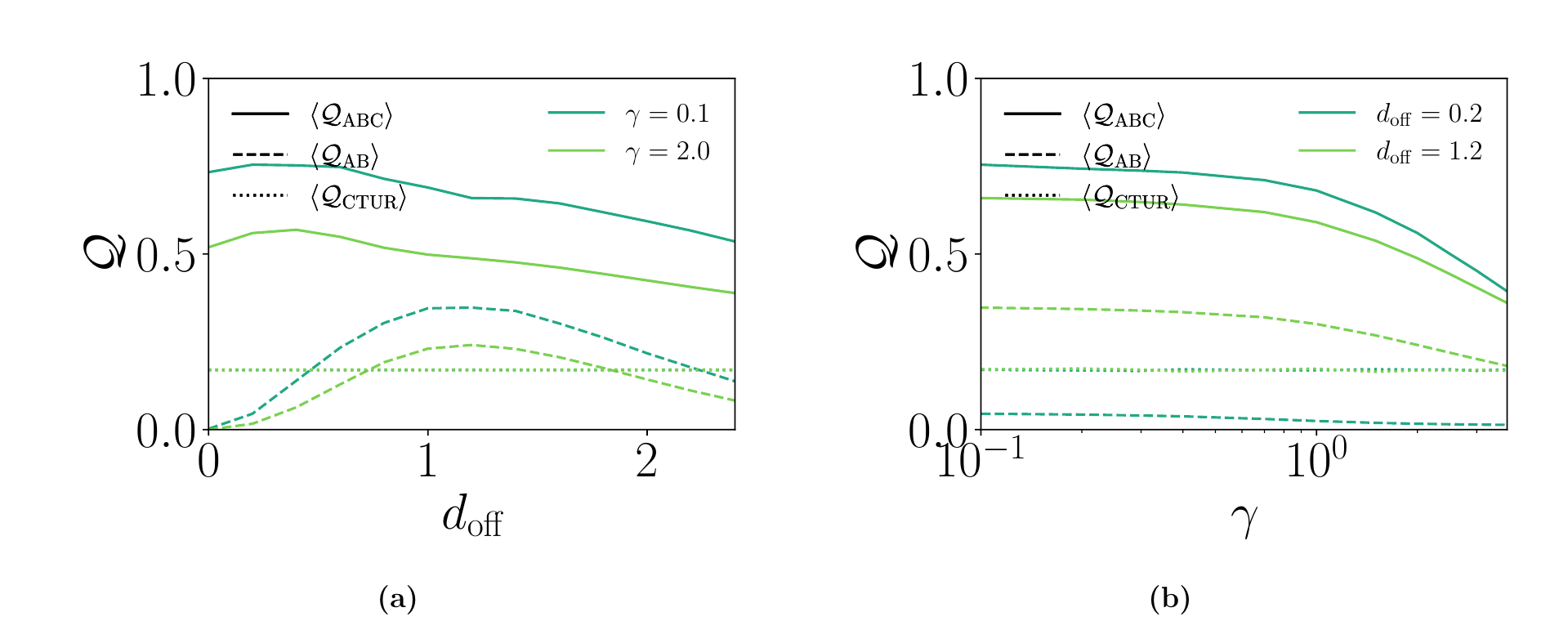}
    \caption{Quality factor of different estimators in the Brownian vortex model. (a) Quality factor as a function of the distance of the center of the two observable regions from the $x$-axis, $d_{\mathrm{off}}$, cf. \Cref{fig:kartoffelwurm} (a). There is an optimal intermediate offset for which the quality factor $\mathcal{Q}_{\text{AB}}$ reaches its maximum. (b) Quality factor as a function of driving strength $\gamma$. All three lower bounds perform better close to equilibrium. Parameters used are $d_{AB}=2$, $r_A=r_B=0.75$, $T=6\times10^7$, $T_{\mathrm{CTUR}}=100$, $\Delta t=10^{-2}$.}
    \label{fig:num_results}
\end{figure*}

We demonstrate our result with numerical simulations of a Brownian vortex in a harmonic trap. The harmonic potential is given by
\begin{equation}
    V(\vec x) = \frac{\kappa}{2}\left(x^2+y^2\right)\,,
    \label{eq:num_pot}
\end{equation}
with the trap strength $\kappa$. The non-conservative driving force reads
\begin{equation}
    \vec f(\vec x) = \gamma\begin{pmatrix} y\\-x
    \end{pmatrix}\,,
    \label{eq:num_force}
\end{equation}
with the driving strength $\gamma$. Further, to keep the model simple, we assume that the diffusion matrix is diagonal
\begin{equation}
    \vec D(\vec x) = D\mathbf{I}\,,
\end{equation}
with the identity matrix $\mathbf{I}$. Since the non-conservative force and the potential are radially symmetric and the diffusion coefficient is homogeneous, the steady-state probability distribution is the equilibrium Boltzmann distribution
\begin{equation}
    \rho(\vec x) \propto \exp\left[-V(\vec x)\right] \,,
\end{equation}
which allows for a straight-forward calculation of the mean entropy production rate of this system. The mean local velocity is given by
\begin{equation}
    \vec v(\vec x) = D\left[-\nabla V(\vec x)+\vec f(\vec x)-\nabla\ln\rho(\vec x)\right] = D\vec f(\vec x)\,.
\end{equation}
Plugging this into \cref{eq:def_sig_nu_D_nu} yields
\begin{equation}
    \sigma =D\gamma^2\left(\langle x^2\rangle+\langle y^2\rangle\right) =\frac{2D\gamma^2}{\kappa}\,.
\end{equation}

We consider two different observation scenarios for our demonstration. In the first one, the particle can be detected if it is in region $A$ and if it is in region $B$ as shown in \Cref{fig:kartoffelwurm}(a). In the second scenario, crossings of the positive $y$-axis and their respective direction can be resolved in addition to observing the regions $A$ and $B$. We denote crossing from right to left as $C_+$ and the time-reversed crossings from left to right as $C_-$. In this scenario, we can then measure the finite-time current across the boundary $C$ in addition to the time spent in $A$ and the time spent in $B$. We can then apply the CTUR \cite{horo17,piet17,dech21} and compare it to our estimator \eqref{eq:main_res} which includes the events $\{A_\pm, B_\pm, C_\pm\}$. 

We simulate long trajectories using the Euler-Maruyama method to obtain empirical frequency distributions $\hat{\nu}_{I\to J}(t_i)$. We do so by counting the number of transitions $N_{I \to J}(t_i)$ between the observable events that take a time $t_i\leq t<t_i+\Delta t$. Since we cannot simulate infinitely long trajectories, we need to discretize the continuous time argument of $\nu_{I\to J}(t_i)$. The empirical histogram of $\nu_{I\to J}(t)$ after a total simulation time $T$ is then given by
\begin{equation}
    \hat{\nu}_{I\to J}(t_i) = N_{I \to J}(t_i)/T\Delta t \,.
\end{equation}
We plug this quantity in \cref{eq:main_res} to obtain the lower bounds $\sigma_{\mathrm{AB}}$ and $\sigma_{\mathrm{ABC}}$ for each observation scenario.

The empirical current is sampled by counting the number of occurrences of $C_+$ and $C_-$ within a time frame $iT_{\mathrm{CTUR}}\leq t<(i+1)T_{\mathrm{CTUR}}$ with $0\ll T_{\mathrm{CTUR}}\ll T$, which we denote $N^i_{C_\pm}$. Along a long trajectory, this gives many samples of the empirical current
\begin{equation}
    \hat{j}^i = \frac{N^i_{C_+}-N^i_{C_-}}{T_{\mathrm{CTUR}}}\,,
    \label{eq:emp_curr}
\end{equation} 
which is the integral of the current density (2) across the positive $y$-axis. Across this line, recrossing events do not contribute to the net difference between the number of jumps in either direction. At the end of the observable boundary, paths starting in the segment $\left[0,\epsilon \right)$ on the y-axis can either recross, or they can cross through the unobserved segment $\left[-\epsilon,0\right)$ and cross through $\left[0,\epsilon \right)$ again after. However, the later case no longer includes recrossing events and is thus a well-defined contribution to the current across the positive $y$-axis.

From these empirical current samples, we can calculate the sample mean $\langle\hat{j}^i\rangle$, the sample variance $\mathrm{var} (\hat{j}^i)$. As in \cite{dech21}, we consider the time-integral of the inverse occupation fraction as state observable $Z$, with the empirical value
\begin{equation}
    \hat{Z}^i = \sum_s \frac{\tau^i_s}{\langle \tau_s\rangle}\,,
\end{equation} 
where $\tau^i_s$ is the time spent in state $s$ during the trajectory sample $i$ and $s\in\{A,B,\text{not detected}\}$, such that $\sum_s \tau^i_s = T_{\mathrm{CTUR}}$. The CTUR is then calculated using the Pearson correlation coefficient 
\begin{equation}
    \chi_{\hat{j},\hat{Z}} = \frac{\mathrm{cov}(\hat{j}^i,\hat{Z}^i)}{\sqrt{\mathrm{var}(\hat{j}^i)\mathrm{var}(\hat{Z}^i)}}\,,
\end{equation} 
as
\begin{equation}
   \sigma_{\mathrm{CTUR}} = \frac{2\langle (\hat{j}^i)^2\rangle}{T_{\mathrm{CTUR}}\mathrm{var}(\hat{j}^i)(1-\chi_{\hat{j},\hat{Z}}^2)}\,.
\end{equation}

For a sensible comparison between $\sigma_{\mathrm{AB}}$, $\sigma_{\mathrm{ABC}}$ and $\sigma_{\mathrm{CTUR}}$, we consider the quality factor 
\begin{equation}
    \mathcal{Q}_m=\frac{\sigma_{m}}{\sigma}\,, \quad m\in\{\mathrm{ABC,AB,CTUR}\}\,,
\end{equation}
which quantifies how much of the true entropy production rate is recovered by each estimator.

We vary the distance of the centers of $A$ and $B$ to the $x$-axis, which we denote $d_{\mathrm{off}}$ and the strength of the non-conservative driving $\gamma$. The results are shown in \Cref{fig:num_results}, where we show the quality factor as a function of $d_{\mathrm{off}}$ and $\gamma$.

In \Cref{fig:num_results} (a), there is an optimal offset $d_{\mathrm{off}}$ for $\mathcal{Q}_{\mathrm{AB}}$. For $d_{\mathrm{off}}=0$, the measured waiting-time distributions are equal due to symmetry. Conversely, for $d_{\mathrm{off}}\to\infty$ the current introduced by the vortex vanishes due to the confining potential. At the optimal offset, this estimator outperforms the CTUR without measuring any time asymmetric observables. Even more remarkably, the observation of only two states contains no cycles such that there are as many transitions $A_+\to B_-$ as $B_+\to A_-$. Thus, the entire inferred entropy production rate stems from the asymmetry in the time a transition $A_+\to B_-$ takes compared to a transition $B_+\to A_-$.

For $\sigma_{\mathrm{ABC}}$, the additional observation of $C_+$ and $C_-$ breaks the symmetry of the system even at $d_{\mathrm{off}}=0$. This lower bound contains the cycles $A_+\to B_-\to B_+\to C_+\to A_-$, $A_+\to C_-\to A_-$, $B_+\to C_+ \to B_-$ and their time reversed counterparts. In this way, $\sigma_{\mathrm{ABC}}$ performs best since it can incorporate the most amount of information. 

In \Cref{fig:num_results} (b) we show the three quality factors as a function of $\gamma$. With increasing driving strength, i.e., further from equilibrium, the quality factor of each lower bound decreases.
\section{Conclusion}
\label{sec:conclusion}
We have presented a novel lower bound on the entropy production rate based on the frequency of transitions of a single particle between coarse-grained regions of space. The strength of our bound is that it includes the full information of the transition-times between these regions and that it does not rely on the observation of Markovian events. As a trade-off, it yields a looser bound than if Markovian events were accessible through a more detailed level of observation, technically arising from the log-sum inequality \eqref{eq:log_sum}. 

In continuous systems, waiting-time distributions between observable, and thus absorbing, boundaries are ill-defined. We have addressed this issue by introducing two kinds of discretization, which shows that relevant functions of these quantities, in particular the frequency of transitions with a certain transition time, remain well-defined in the continuum limit. In \cite{meyb24}, this issue was circumvented by considering composite events, i.e., consecutive crossings of certain points, as renewal events. In such a scheme, the particle is not placed on an absorbing boundary when calculating the waiting-time distribution. However, the approach in \cite{meyb24} could only be applied in one dimension. Thus, we think that transition frequencies, rather than waiting-time distributions, might be of interest in other systems with different kinds of Markovian events and lead to further lower bounds on entropy production. 

The log-sum inequality \eqref{eq:log_sum} holds for any many-to-one coarse-graining of the underlying Markovian events that leads to even or odd transition classes such as the one presented here and in \cite{erte24}. Further, the factorization \cref{eq:factorization} is generic and thus the resulting expressions $\sigpw$ and $\sigloc$ also appear quite generally for such coarse-graining schemes. The challenge is to meaningfully identify $\sigloc$ as a lower bound to entropy production for a given type of Markovian event. For blurred transitions in Markov networks this has been achieved in Ref. \cite{erte24} which we have generalized to blurred infinitesimal transitions in overdamped Langevin dynamics. We expect that identifying $\sigloc$ as a lower bound to entropy production is possible in more coarse-graining schemes and offers a promising path for further research. 

Finally, it would be interesting to study this coarse-graining scheme for multi-particle systems and for underdamped dynamics.

\section*{Acknowledgements}
We thank Alexander M. Maier for insightful discussions.
 
\input{main.bbl}
\end{document}

%% file: main.bbl
%